\documentclass[prl,showpacs,twocolumn,superscriptaddress]{revtex4-1}

\usepackage{amsmath,graphicx,latexsym,bm}

\begin{document}

\title{First-principles modeling of electrostatically doped 
perovskite systems}

\author{Massimiliano Stengel}
\affiliation{Institut de Ci\`encia de Materials de Barcelona (ICMAB-CSIC), Campus UAB, 08193 Bellaterra, Spain
}
\email{mstengel@icmab.es}
\date{\today}

\begin{abstract} 
Macroscopically, confined electron gases at polar oxide interfaces 
are rationalized within the simple ``polar catastrophe'' model.
%
At the microscopic level, however, many other effects such as
electric fields, structural distortions and quantum-mechanical 
interactions enter into play.
%
Here we show how to bridge the gap between these two length scales, by 
combining the accuracy of first-principles methods with the conceptual
simplicity of model Hamiltonian approaches.
To demonstrate our strategy, we address the equilibrium distribution of 
the compensating free carriers at polar LaAlO$_3$/SrTiO$_3$ interfaces.
Remarkably, a model including only calculated bulk properties of SrTiO$_3$ 
and no adjustable parameters accurately reproduces our full first-principles results.
Our strategy provides a unified description of charge compensation
mechanisms in SrTiO$_3$-based systems.
\end{abstract}


\maketitle


The unusual behavior of the (001) LaAlO$_3$/SrTiO$_3$ interface is commonly 
understood in terms of the ``polar catastrophe'' (PC) model.~\cite{Ohtomo-04}
By stacking charged (LaO)$^+$ and (AlO$_2$)$^-$ layers on top of the charge-neutral
TiO$_2$ and SrO layers, one obtains a net interface charge density of 
$\sigma_{PC}=+e/2S$, where $S$ is the unit-cell cross section.
This produces a diverging electrostatic energy, unless $\sigma_{PC}$ is
neutralized by an external free charge, which would explain the
appearence of confined mobile carriers at this interface.

This model, while appealing, misses many important effects, that are crucial 
for a realistic description of the interface.
For example, it was shown that strong polar distortions in 
LaAlO$_3$~\cite{Pentcheva-09} and/or in SrTiO$_3$~\cite{Stengel-09-quack.1,Bristowe-09} 
partially screen the excess charge, delaying the onset of
metallicity far beyond what the PC arguments would predict.
Next, it was shown that H adsorbates~\cite{Son-10} or oxygen 
vacancies~\cite{Cen-09} at the open LAO surface can profoundly 
alter the sheet density of free carriers. Moreover, reversible 
metal-insulator transitions can also be induced upon
application of an external bias~\cite{Cen-08,Thiel-06}.
Both effects go clearly beyond the oversimplified PC description.
Finally, the PC model cannot predict truly \emph{microscopic} 
properties of the system, such as the spatial decay and confinement
of the free electrons near the interface. In an attempt to answer 
these latter crucial questions, various quantum-mechanical 
explanations were proposed~\cite{Chen/Kolpak-10,Janicka-09}, but their
relative importance, especially in relationship to the macroscopic 
electrostatics arguments, is unclear.

\begin{figure}
\begin{center}
\includegraphics[width=3.0in,clip]{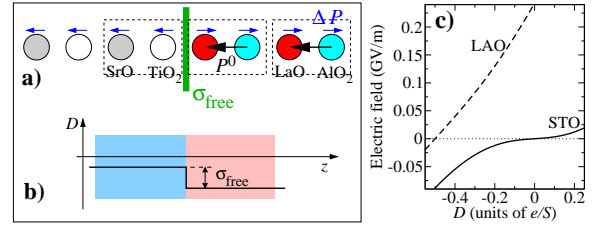} 
\end{center}
\caption{a) Scheme of the TiO$_2$:LaO interface, showing
the built-in ($P^0$, black arrows) and induced ($\Delta P$,
blue arrows) polarization, and the free charge $\sigma_{\rm free}$. 
b) Electric displacement as a function of $z$, illustrating Eq.~\ref{eqsigma}.
c) Calculated internal electric fields $\mathcal{E}$ in bulk STO and LAO as a 
function of $D$. The polar nature of the interface stems from the impossibility
of finding a common value of $D$ for which $\mathcal{E}_{\rm LAO}$ and 
$\mathcal{E}_{\rm STO}$ are both zero.
\label{fig0} }
\end{figure}

A route towards bridging the gap between classical electrostatics
and quantum theory in LAO/STO and related systems was proposed 
in Ref.~\onlinecite{Stengel-09-quack.1}. 
The strategy is based on the ``formal''~\cite{ferro:2007} definition 
of the polarization, $P$, in quantum-mechanical systems, which has
a simple classical interpretation in terms of a point-charge model
(see Fig.~\ref{fig0}).
%
The dipole moment of an individual (LaO)-(AlO$_2$) unit is $d=-ea/2$ 
[black arrows in Fig.~\ref{fig0}(b)], where $a$ is the out-of-plane 
lattice parameter and $e$ is the (positive) electronic charge. This 
corresponds to a ``built-in'' polarization $P^0_{\rm LAO}=-e/2S$,
where $S$ is the cell surface. Conversely, $P^0_{\rm STO}=0$ because 
the STO layers are formally charge-neutral.
Note that there is no left-over ionic charge at the interface -- we have 
reinterpreted $\sigma_{PC}$, as a surface density of \emph{bound} charge, 
that arises because of a discontinuity in $P$.


Depending on the electrical boundary conditions, macroscopic electric 
fields, respectively $\mathcal{E}_{\rm LAO}$ and $\mathcal{E}_{\rm STO}$, 
will be present on one or either side of the junction.
The electric fields will in turn perturb the individual LAO and STO layers,
producing an ``induced'' polarization that we call $\Delta P_{\rm LAO}$ 
and $\Delta P_{\rm STO}$. 
If we now define the total polarization as $P = P^0 + \Delta P$,
and the electric displacement as $D=\epsilon_0 \mathcal{E} + P$,
an \emph{exact} relationship follows,~\cite{Stengel-09-quack.1}
\begin{equation}
D_{\rm LAO} - D_{\rm STO} = \sigma_{\rm free},
\label{eqsigma}
\end{equation}
where $\sigma_{\rm free}$ is a surface density of ``free'' charge
confined to the interface region.
Eq.~\ref{eqsigma} generalizes the PC model by taking rigorously 
into account the effect of polar distortions ($\Delta P$ is
implicitly contained in $D$), external biases [$\mathcal{E}(D)$ 
is a bulk property of either material, and is a unique function 
of $D$, see Fig.~\ref{fig0}(c)] and charged species adsorbed on 
the far-away surfaces (the flux of $D$ corresponds to the surface 
charge density).
By appropriately choosing the two independent parameters $D_{\rm LAO}$
and $D_{\rm STO}$ one can therefore describe the local properties of
an ideal interface within arbitrary boundary conditions, encompassing
virtually all theoretical approaches (stoichiometric or non-stochiometric
superlattices and various flavors of slab geometries) that were
used so far in the literature.~\cite{Pentcheva-10,Sohrab-10}
One does not need to worry about the specific mechanisms and/or
supercell geometries that determine a certain equilibrium value of 
$D_{\rm LAO}$ and $D_{\rm STO}$, as long as the interface can be 
thought as isolated (say, separated by at least two or three unit 
cells of LAO and STO on either side).
%
%

To work our way towards the microscopics, it is now tempting to take the 
analogy to macroscopic Maxwell equations one step further, and write
\begin{equation}
\frac{dD(z)}{dz} = \rho_{\rm free}(z).
\label{eqdd}
\end{equation}
Here $\rho_{\rm free}(z)$ is the spatially resolved planar average
of free carriers, whose integral along $z$ yields $\sigma_{\rm free}$.
It is easy to verify that Eq.~\ref{eqdd} is consistent with Eq.~\ref{eqsigma}.
Here one runs into trouble, however, as one needs to establish a truly
\emph{microscopic} definition of both $D(z)$ and $\rho_{\rm free}$. (Note that 
this is not necessary at the level of Eq.~\ref{eqsigma}, which deals only
with macroscopic quantities.)
This is a nontrivial issue in a typical metal, where the polarization
(and hence $D$) is ill-defined. Furthermore, $\rho_{\rm free}$ is microscopically 
difficult to identify, as the bands corresponding to the conduction electrons are
generally entangled with lower-lying bound states.
In a doped oxide or semiconductor, however, the valence and conduction
bands usually preserve their identity, i.e. a well-defined energy gap 
persists between conduction-band and valence-band states. 
This naturally leads to a definition of $\rho_{\rm free}$ based 
on the overall density of the partially occupied states near the Fermi level.
The remainder is an \emph{integer} number of electrons that we identify as
bound charges. We use these latter orbitals to define a layer-resolved
electric displacement based on a Wannier decomposition of the 
polarization,~\cite{xifan_lp,Stengel-09-quack.1} in analogy with
standard insulators.

\begin{figure}
\begin{center}
\includegraphics[width=1.8in,clip]{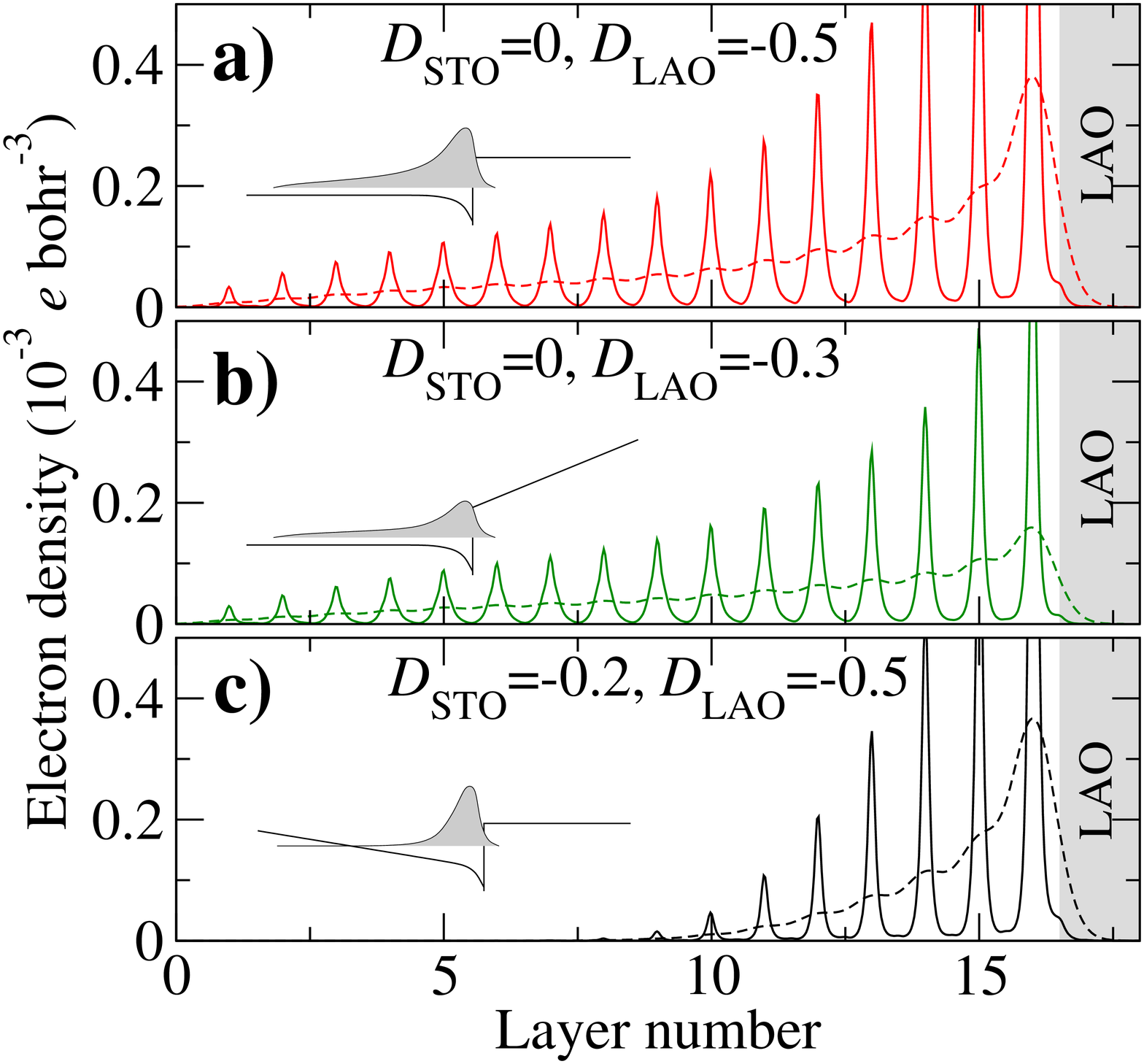} 
\includegraphics[width=1.5in,clip]{fig2b.eps} 
\end{center}
\caption{Conduction charge (a-c) and local electric displacement (d) at the 
          SrTiO$_3$/LaAlO$_3$ interface. In (a-c) the solid curves are 
          $\rho_{\rm free}(z)$ and the dashed curves are the nanosmoothed 
          $\bar{\rho}_{\rm free}(z)$. Insets show an approximate diagram
          of the effective local potential.
          In (d) the symbols show the Wannier-based local electric
          displacement computed from the bound charges. The curves represent
          $\int_{\-\infty}^z \bar{\rho}_{\rm free}(t) dt + D_{\rm STO}$.
          \label{fig1} }
\end{figure}

We are now ready to verify Eq.~\ref{eqdd} directly on our first-principles 
calculations.
To provide a representative number of test cases, we study three combinations 
of $D_{\rm STO}$ and $D_{\rm LAO}$, which are summarized in the insets of Figure~\ref{fig1}.
Case (a) corresponds to full compensation, e.g. that of a thick LAO overlayer
on a thick STO substrate. Case (b) corresponds to partial compensation, which 
can occur at intermediate LAO thicknesses~\cite{Son_et_al:2009}, or in the case of an
electrical bias applied between the electron gas and an electrode deposited
on the free surface.
Case (c) physically corresponds to a ``back-gating'' regime, where an electrical
bias is applied between the electron gas and an electrode placed at the other end
of the STO substrate.
In practice, we use slab geometries of the type vacuum/(SrTiO$_3$)$_n$/(LaAlO$_3$)$_m$/vacuum 
(we use $n=16$ and $m=3$ in our calculations), where the boundary conditions on $D$
are enforced as explained in the Supplementary Information. (All the other relevant
computational parameters are also described there.)
In Fig.~\ref{fig1} we show the relaxed $\rho_{\rm free}(z)$ (a-c) and the 
layer-by-layer (locally averaged) electric displacement $D_l$ (d) for each 
combination.
In Fig.~\ref{fig1}(d) we also plot three curves that we constructed
by numerically integrating the nanosmoothed charge densities, $\bar \rho_{\rm free}(z)$.
[We are therefore verifying the integral version of Eq.~\ref{eqdd}, 
$D(z) = \int_{-\infty}^z \bar{\rho}_{\rm free}(t) dt + D(-\infty)$.]
The matching is excellent in all cases, demonstrating the high accuracy of 
Eq.~\ref{eqdd}.
Note that in one of the examples (case c) $\rho_{\rm free}$ decays to zero 
relatively fast when moving away from the interface, while it spreads over the 
whole volume of the SrTiO$_3$ film in the other two cases. This is due to the fact 
that in c) the asymptotic electric field in STO is not zero, but equal to 
$\mathcal{E}(D=-0.2)\sim$-12 MV/m [see Fig.~\ref{fig0}(c)]. This produces a 
confining wedge potential, that limits the spread of $\rho_{\rm free}$. 
Conversely, in (a) and (b), $\mathcal{E}$ vanishes at $z \rightarrow -\infty$,
and the outermost electrons are only loosely bound.

Eq.~\ref{eqdd} is an important result, in that it establishes a direct, 
virtually exact relationship between the density of compensating carriers and 
the local polarization in LAO and STO. 
This answers pressing experimental questions concerning precisely this point,
as polar distortions in SrTiO$_3$ were recently observed.~\cite{Ramesh-10}
%
%
This also has profound implications over the theoretical understanding
of electron confinement in this system, as we shall demonstrate in the following.

\begin{figure}
\begin{center}
\includegraphics[width=3.0in,clip]{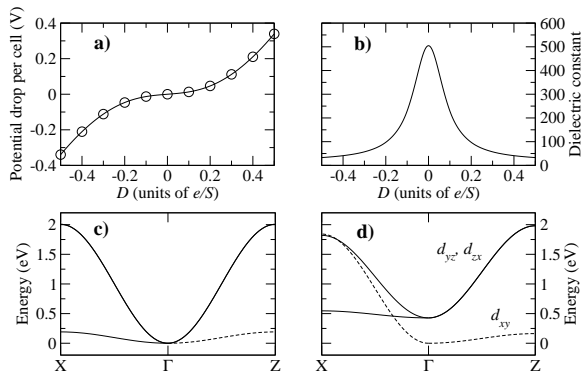} 
\end{center}
\caption{Bulk properties of SrTiO$_3$ used in the tight-binding model. Upper panels:
          internal potential (a) and dielectric constant (b) as a function of $D$.
          Lower panels: $t_{2g}$ conduction band structure for $D=0$ (c) and
          $D=-e/2S$ (d). \label{fig2} }
\end{figure}

Essentially, the equilibrium distribution of the conduction charge
is determined by two competing effects. One is the electrostatic 
energy, that tends to localize the electrons as close to the interface 
as possible. The strength of the attraction will depend on the static 
dielectric constant of the underlying insulator.
The other is the quantum-mechanical kinetic energy of the electrons. This 
will tend to spread the electrons in space, with a strength that depends 
on the band dispersion. 
To see whether, and to what extent, the large polar distortions in 
STO affect these competing driving forces, we performed
calculations of bulk SrTiO$_3$ as a function of the electric 
displacement~\cite{fixedd}, by covering the range of $D_{\rm STO}$ 
values that are relevant for the LAO/STO system. 
For each value of $D$ we extract the built-in electric field, the total 
internal energy and the relevant parameters of the lower part of the 
conduction band. These are the tight-binding hopping integrals between 
Ti-derived orbitals with $t_{2g}$ symmetry ($d_{xy}$, $d_{xz}$ and $d_{yz}$).
As the $t_{2g}$ orbitals are fairly well localized in space, it is sufficient 
for the present study to consider only the first three shells of nearest-neighbors
Ti sites. 

In Fig.~\ref{fig2}(a) we show the electric field as a function of 
the displacement field $D$. Note the strong nonlinearity, which 
is evident in the plot of the dielectric constant as a function of
$D$ [Fig.~\ref{fig2}(b)].
In Fig.~\ref{fig2}(c) we show the band structure as it results from
the third-neighbor Hamiltonian, for the centrosymmetric cubic state
at $D=0$. 
Note the symmetry of the bands, which are characterized by a three-fold 
degeneracy at $\Gamma$.
%
A polarization [the extreme case $D=-e/2S$ is shown in Fig.~\ref{fig2}(d)] 
lifts this degeneracy, by producing a strong splitting at $\Gamma$ between 
the degenerate $d_{xz}/d_{yz}$ orbitals and the $d_{xy}$ orbital.
This splitting is dominated by the strong reduction in 
the $d_{xz}/d_{yz}$ bandwidth along the $\Gamma \rightarrow X$ and $\Gamma \rightarrow Z$ 
directions -- the corresponding hopping terms are reduced by as much as 
30\% and 25\%, respectively.
Polarization-related changes in other matrix elements appear to be less 
pronounced.

We shall now use these data to develop a quantitative model of the 
equilibrium distribution $\rho_{\rm free}(z)$.
We make a rather bold assumption here, and state that the role played by 
the LaAlO$_3$ overlayer in determining $\rho_{\rm free}(z)$ is marginal, 
except for two crucial effects: i) it confines the conduction electrons to 
the STO side, and ii) it defines the electrical boundary conditions through 
the value of $D_{\rm LAO}$.
Based on this \emph{Ansatz}, we represent the LAO/STO interface systems
discussed in the previous paragraphs as \emph{pure} STO slabs, periodic in
plane and $n$-layer thick, where the boundary values of the electric 
displacement field at the two surfaces are set to $D_{\rm STO}$ and 
$D_{\rm LAO}$.
To each Ti site $l$ we assign three orbitals of $t_{2g}$ symmetry, and
a charge density $\rho_l$. The charge density defines the local value
of the electric displacement $D_l$ through Eq.~\ref{eqdd}. The Hamiltonian 
matrix elements are defined by the electrostatic potential $V_l$ [calculated from
$D_l$ using the bulk $V_{\rm STO}(D)$ of Fig.~\ref{fig2}(a)], which 
rigidly shifts the on-site terms, and by the $D_l$-dependent hopping 
parameters that we interpolate from the bulk SrTiO$_3$ data. Upon 
diagonalization we obtain the wavefunctions, that self-consistently
determine $\rho_l$ within the constraint Eq.~\ref{eqsigma}. 
%

\begin{figure}
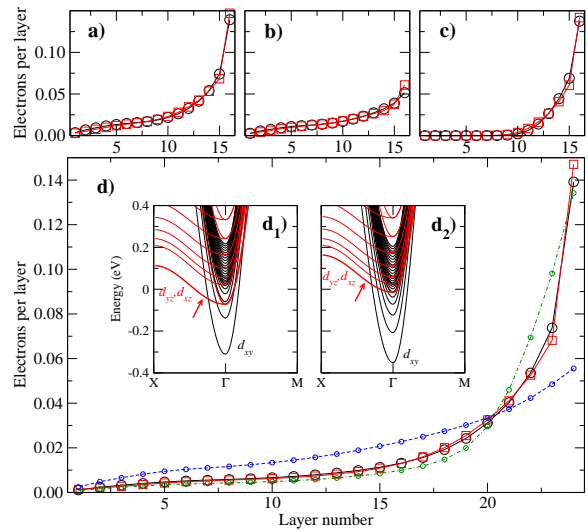

\begin{center}
\includegraphics[width=3.0in,clip]{fig4a.eps} \\
\includegraphics[width=3.0in,clip]{fig4b.eps} 
\end{center}
\caption{Electron confinement and spatial distribution. Red curves with square symbols are 
          the first-principles results, black curves (circles) are the results 
          of the model. Parameters are the same as in Fig.~\ref{fig1} (a-c). 
          (d) shows a thicker 24-cell STO film. The dashed (blue) and dot-dashed (green) 
          curves were obtained by artificially 
          altering selected features of the model (see text). Insets: tight-binding band 
          structures corresponding to
          the green (d$_1$) and black (d$_2$) curves in (d). The arrow
          indicates the lowest $d_{yz}/d_{zx}$ state.
          \label{fig3} }
\end{figure}

In Fig.~\ref{fig3} we compare the results of the model to the first-principles 
simulations discussed earlier. We include in the comparison a fourth simulation 
that we did for case $D_{\rm STO}=0,D_{\rm LAO}=-e/2S$, but with a thicker STO 
layer ($n=24$).
The agreement is remarkably good. This indicates that the polarization-dependent 
bulk properties of SrTiO$_3$, together with the boundary values of $D$, are sufficient 
to explain the distribution of conduction charge in this system. 
This suggests that the interaction between Ti- and La-derived orbitals is
not an essential factor in determining electron confinement, contrary to
the conclusions of Ref.~\onlinecite{Chen/Kolpak-10}.
Binding of the electrons to the interface is indeed guaranteed by 
Eq.~\ref{eqdd}.

Now that we have a reliable model we can directly quantify the impact of each
specific STO bulk property on the distribution of $\rho_{\rm free}$. 
First, if we neglect the non-linearity in the dielectric permittivity $\epsilon_{\rm STO}(D)$, 
and instead use a constant $\epsilon_{\rm STO}(D)=\epsilon_{\rm STO}(0)\sim 500$  
we obtain a much broader distribution [blue curve in Fig.~\ref{fig3}(d)].
%
This indicates that the carrier distribution is strongly sensitive to 
the dielectric properties of bulk STO; this seems to be an accepted
fact in the experimental community,~\cite{Copie-09} but has 
received surprisingly little attention in earlier \emph{ab-initio} studies. 
Second, if we suppress the $D$-dependence of the STO band structure,
and use the $D=0$ $t_{2g}$ Hamiltonian throughout the film, we obtain [green 
curve in Fig.~\ref{fig3}(d)] an excessive accummulation of charge 
in the near-interface region.
This effect can be understood by comparing the self-consistent bandstructures
of the original [Fig.~\ref{fig3}(d$_2$)] and the ``$t_{2g}(D=0)$'' 
[Fig.~\ref{fig3}(d$_1$)] tight-binding models.
In both cases there is a strong splitting at $\Gamma$ between the $d_{xy}$
and $d_{xz}$/$d_{yz}$ bands, in agreement with the findings of 
Refs.~\onlinecite{Popovic-08,Son_et_al:2009,Sohrab-10}. 
In Fig.~\ref{fig3}(d$_1$), however, this splitting is only induced by \emph{confinement}
effects due to the wedge-like electrostatic potential near the 
interface~\cite{Gervasi-10}. 
%
The polarization-induced perturbations in the STO $t_{2g}$ bands 
[Fig.~\ref{fig2}(d)] significantly enhance such a splitting [Fig.~\ref{fig3}(d$_2$)] 
and shift the $d_{xz}$/$d_{yz}$ bands further up in energy. (The effect is 
strongest on the lowest $d_{xz}$/$d_{yz}$ band, marked with a red 
arrow in the figure).
This upshift, in turn, pushes the weight of the $d_{xz}$/$d_{yz}$ electrons 
away from the LAO interface, which explains the difference between the 
respective electron distributions [green and black curves in Fig.~\ref{fig3}(d)].
%
%
%
%
%
Note that an analogous $t_{2g}$ splitting was experimentally observed in 
LAO/STO,~\cite{Salluzzo-09} and theoretically also discussed in the context
of the closely related LaTiO$_3$/SrTiO$_3$ system.~\cite{Okamoto-06}
%

%
%
%

The tight-binding method used here has clear points of contact 
with the strategy of Refs.~\onlinecite{Okamoto-04,Okamoto-06}. 
However, in our approach there is a crucial innovation. 
%
Here, at difference with Ref.~\onlinecite{Okamoto-06}, we
extract all the ingredients of the model from \emph{bulk} calculation 
of pure insulating SrTiO$_3$, without including any adjustable parameter.
%
This 
forces us to build a \emph{universal} and transferable model, which 
can be readily applied to essentially any situation involving 
electrostatic doping of SrTiO$_3$, and is not restricted to the specifics 
of the LaAlO$_3$/SrTiO$_3$ interface. 
For example, our strategy could be readily used, with little modifications,
to interpret the recent findings of electron gases at the bare SrTiO$_3$
surface.~\cite{Gervasi-10}
More importantly, our model could be readily extended to account for
other physical ingredients not considered here, e.g. strong 
correlations~\cite{Pentcheva-10} and strain effects;~\cite{Bark-10}
all we need to do is to refine the theoretical description of \emph{bulk}
STO that we take as input.
%
%
This is an enormous advantage, both conceptually (the model is based
on few parameters that are easy to interpret) and practically (the 
tight-binding model is several orders of magnitude more efficient than a
full first-principles calculation).
%
%
More generally, our results open exciting new avenues for the study of confined
electron gases in oxide systems, with optimal accuracy and dramatically reduced 
computational cost.

I wish to thank G. Herranz, J. \'I\~niguez and Ph. Ghosez for their
careful reading of the manuscript. This work was supported by DGI-Spain 
(Grants No. MAT2010-18113 and No. CSD2007-00041) and EC-FP7 
(Grant No. NMP3-SL-2009-228989). Computing time was kindly provided by
BSC-RES and CESGA.

\bibliography{max}

\end{document}